\documentclass[aps,prl,nofootinbib,superscriptaddress,onecolumn,notitlepage,preprintnumbers]{revtex4-1}

\usepackage[english]{babel}

\usepackage{amsmath,graphicx,verbatim,epsfig}
\usepackage{amsmath,graphicx}
\usepackage{color}
\usepackage{datetime}
\usepackage{slashed}

\usepackage{subfig}
\usepackage{mathtools}

\usepackage{afterpage}
\usepackage[normalem]{ulem} 

\newcommand{\vek}[1] {\boldsymbol{#1}}

\newcommand{\y}{\vek{y}}

\newcommand{\nn}{\nonumber}

\allowdisplaybreaks[3]

\begin{document}

\preprint{MITP/18-090}
\preprint{Nikhef 2018-047}


\title{A spin on same-sign W-boson pair production}

\author{Sabrina Cotogno}
\email{scotogno@nikhef.nl}
\affiliation{Nikhef, Science Park 105, 1098 XG Amsterdam, The Netherlands}
\affiliation{Department of Physics and Astronomy, VU
  University Amsterdam, De Boelelaan 1081, 1081 HV Amsterdam, The
  Netherlands}
  
\author{Tomas Kasemets}
\email{kasemets@uni-mainz.de}
\affiliation{PRISMA Cluster of Excellence \& Mainz Institute for Theoretical Physics\\ Johannes Gutenberg University, 55099 Mainz, Germany}

\author{Miroslav Myska}
\email{miroslav.myska@fjfi.cvut.cz}
\affiliation{FNSPE, Czech Technical University in Prague, Brehova 7, 115 19 Prague, Czech Republic}


\begin{abstract}
We demonstrate that the LHC will be sensitive to quantum correlations between two quarks inside the proton. Same-sign W-boson pair production is the most promising channel for clear measurements of double parton scattering. The left-handed nature of the coupling between quarks and W-bosons makes it a prime probe to measure parton spin correlations. We perform a detailed analysis of double parton scattering, including relevant backgrounds. The analysis reveals that measurements comparing the rate at which two muons from W boson decays are produced in the same compared to opposite hemispheres are especially sensitive to spin correlations between two quarks inside the proton. We provide estimates of the significance of the measurements as a function of the integrated luminosity. 
\end{abstract}

\maketitle

\section{Introduction}
The LHC will allow us to have a precise and deep enough look inside the proton to reveal how the quantum properties of two quarks are interconnected. That correlations between the quantum numbers of quarks and gluons inside the proton can in principle be probed through double parton scattering (DPS) was realized in a series of papers by Mekhfi \cite{Mekhfi:1985dv,Mekhfi:1983az}. Although this was an exciting discovery, the papers were somewhat ahead of time and attracted only mild attention. The discussion of quantum correlations, and in particular spin correlations in DPS came back to life in the early days of the LHC \cite{Manohar:2012jr,Diehl:2011yj,Diehl:2011tt}. Although relatively widespread, these were  to a large extent still exercises with little impact on experimental searches and observables. Even for the only case so far where it has been shown that spin correlations have a large influence on the cross section \cite{Echevarria:2015ufa}, no clear observable for their detection was found. The reason was that the spin correlations had a large impact on the (unknown) size of the cross section, but only milder effects on the distributions of final state particles. For a summary of the theoretical status of DPS and correlations between two partons we refer to Ref.~\cite{Diehl:2017wew,Kasemets:2017vyh,Blok:2017alw,Treleani:2018dbg,Blok:2013bpa} and references therein. In short, many of the elements required in a proof of factorization of DPS have been established, with a consistent separation between single and double parton scattering and with double parton distributions (DPDs) describing the properties of the two partons inside the proton \cite{Buffing:2017mqm,Diehl:2017kgu,Diehl:2015bca}. 

Same-sign W-pair (SSW) production is one of the most studied DPS processes \cite{Cao:2017bcb,Ceccopieri:2017oqe,Luszczak:2014mta,Golec-Biernat:2014nsa,dEnterria:2012jam,Myska:2013duq,Myska:2012dj,Gaunt:2010pi,Kulesza:1999zh}. In particular Ref.~\cite{Ceccopieri:2017oqe}  investigated the effect of kinematical correlations between the partons. The contribution to the signal from single parton scattering is suppressed by additional couplings and produces signatures which makes the two relatively easy to separate experimentally. This implies that the parton splitting mixing single and double parton scattering is suppressed \cite{Diehl:2017kgu,Diehl:2017wew}. LHC is now reaching integrated luminosities large enough to start probing the SSW process and recently first experimental observations or indications of DPS in the SSW final state have been found \cite{CMS:2017jwx,Sirunyan:2017hlu}.

Since the W-boson only couples to left-handed (right-handed) quarks (antiquarks), the polarized DPDs describing the spin correlations between two partons enter linearly into the cross section \cite{Kasemets:2012pr}. This leads to especially large effects on the cross section from spin correlations. In this letter, we demonstrate the impact which spin correlations can have on the distribution and rate of leptons from the W-boson decays. We identify observables with large sensitivity to the spin correlations, taking relevant backgrounds into consideration. We give estimates of the integrated luminosity necessary for measurements to probe the phase space for spin correlations and possibly make the first determinations of quantum correlations between the proton constituents. Further details, including the effects of other types of interparton correlations, a broader range of observables and with different selection criteria, will be discussed in detail in a forthcoming publication \cite{Cotogno:2018xxx}. 

\section{Measuring interparton spin-correlations}
The spin correlations between two partons inside a proton are quantified by polarized DPDs. The SSW process is sensitive to longitudinally polarized quarks (and antiquarks) and the corresponding DPDs describe (at leading order) the difference between the probabilities of finding the two quarks with helicities aligned rather than anti-aligned. These distributions have never been experimentally probed, and are therefore largely unknown. The information that we do have on them, are either thanks to positivity bounds \cite{Diehl:2013mla}, giving upper limits on the polarized distributions, or model calculations \cite{Kasemets:2016nio,Rinaldi:2014ddl,Chang:2012nw}. The correlations decrease with double DGLAP evolution, but  the longitudinally polarized distributions relevant in SSW production can remain large up to high scales (e.g. around 25\% of the unpolarized for the up-antidown distribution at $x_1=x_2=0.01$) \cite{Diehl:2014vaa}.

The purpose of this letter, is to test if and when experimental measurements can provide valuable information about these polarized distributions and thereby access non-trivial information on quantum correlations between proton constituents. To answer this question, we will examine the situations where the bounds are saturated at a low input scale $Q_0$ in such a way that they produce maximal effects of polarization on the cross section. This gives us the relation between the longitudinally polarized and unpolarized distributions for quarks and antiquarks as
\begin{align}\label{eq:polmod}
f_{\Delta a\Delta  b} (x_1, x_2, \y; Q_0) & = (-1)^n f_{a b} (x_1, x_2, \y;Q_0)  \,.
\end{align}
The subscripts $\Delta a$ ($a$) denote longitudinally polarized (unpolarized) partons, $x_i$ is the momentum fraction of parton $i$ and $\y$ is the transverse distance between the two partons. The exponent reads $n=1$ if $a$ and $b$ are both quarks or both anti-quarks and otherwise $n=2$. The scale $Q_0$ should be a low scale, chosen around the scales where perturbative calculations start to be valid. The reason is that once the bounds are saturated at a low scale $Q_0$, they will be satisfied at all larger scales, but typically be violated if perturbation theory is used to evolve them down to even lower scales. We will make this ansatz at the low scale of $Q_0 = 1$~GeV.

Even the unpolarized DPDs are (at best) poorly known. There are a larger number of model calculations compared to the polarized distributions \cite{Rinaldi:2018zng,Rinaldi:2016mlk,Kasemets:2016nio,Broniowski:2016trx,Broniowski:2013xba,Rinaldi:2014ddl,Rinaldi:2013vpa,Chang:2012nw}. Additional constraints on the unpolarized distributions come from sum rules valid after integrating a single DPD over the transverse distance $\y$ \cite{Plossl:2017wjw,Ceccopieri:2014ufa,Blok:2013bpa,Gaunt:2009re}. A common base-line ansatz is to use the approximation that the two partons are uncorrelated, and that there is no significant interdependence between the parton types or momentum fractions and the transverse distance $\y$. We will adopt this scenario throughout this letter, as it allows us to focus on the spin-correlations. Modifying these assumptions, for example by allowing for correlations between the longitudinal momentum fractions, affects the precise numbers but does not alter our conclusions. More details on other types of parton correlations, and other extensions of this simple ansatz, will be the subject of a forthcoming publication \cite{Cotogno:2018xxx}.  The ansatz gives us unpolarized DPDs at a low scale equal to
\begin{align}
	f_{a b} (x_1, x_2, \y; Q_0) & = f_a(x_1;Q_0) f_b(x_2;Q_0) G(\y;Q_0) \,.
\end{align}
The separation of the $\y$ dependence allows us to express the $\y$-dependence at the cross section level in terms of $\sigma_{\text{eff}}=1/\int d^2 \y G(\y;m_W) G(\y;m_W)$ which we will set equal to $15$~mb. This value is in the range extracted by the CMS collaboration in SSW production \cite{CMS:2017jwx}. However, one should be careful not to over-interpret this quantity. The reason for this is, among other things, that the extraction makes assumptions on the shape of the DPS cross section in variables which can be changed by, for example, the spin correlations. Both unpolarized and polarized DPDs are evolved from the initial scale $Q_0$ to the scale of the W-boson mass with (unpolarized and polarized) double DGLAP evolution (without the $1\rightarrow2$ splitting), see e.g. \cite{Diehl:2013mla}.

The total cross section, as well as the distributions of the final state leptons, strongly depend on the polarized distribution. The total DPS cross section varies by up to around 30 percent depending on the model of the polarized distributions. The cross section for double W, including spin correlations, is given in equation (3.5) of reference \cite{Kasemets:2012pr}. The polarized contributions, due to the change in angular momentum when the partons have their helicities aligned rather than anti-aligned, have a direct impact on the rapidity distributions of the muons. 

The SPS contribution to the SSW final state produces additional particles, i.e. $pp \rightarrow W^+ W^+ jj$ with two forward jets. Cuts on these jets effectively remove this background. SSW production is therefore often praised for its background free nature. However, this is a truth with modification, and significant work is necessary to produce a background free DPS sample for the process, see e.g. \cite{CMS:2017jwx,Sirunyan:2017hlu,Gaunt:2010pi}. Main backgrounds are coming from the WZ final state (where one of the muons from the $Z$ boson escapes detection) and QCD, in particular $t\bar{t}$ production. The top quark decaying into a $b$ quark will produce a $\mu^+$, and a second same sign muon can be produced by subsequent B meson decay. The $t\bar{t}$ cross section is, in the absence of cuts etc., many orders of magnitude larger than the DPS signal, but there are a number of effective ways to get this background under control.

We aim to produce a DPS signal which is as pure as possible, and therefore impose restrictive cuts. We analyze particle level final states obtained using a combination of Monte Carlo generators in order to include the effects of the Underlying Event modeling. The DPS signal events come from Herwig 7.1.2. \cite{Bellm:2015jjp,Bahr:2008pv}, where we apply a robust reweighing procedure to mimic our spin correlations at parton level. Herwig is also used for the $t\bar{t}$ production. Diboson samples were obtained via a combination of MadGraph5\_aMC@NLO at LO \cite{Alwall:2014hca} and Pythia 8.2.2.3 \cite{Sjostrand:2014zea,Sjostrand:2006za}. Jets are defined using the default anti-$k_t$ algorithm of FastJet \cite{Cacciari:2011ma} with a pseudo-radius $R$ = 0.4. Our kinematical cuts are 
\begin{align}\label{eq:sel}
	|\eta_i| < 2.4 \; , \quad 25~\text{GeV} < k_{T}^\text{lead} < 50~\text{GeV} \,, \quad 15~\text{GeV} &< 		k_{T}^\text{subl} < 40~\text{GeV}\,, \quad
	k_{T}^{\mu_3} < 5~\text{GeV} \,, \quad \slashed{E}_T > 20~\text{GeV} \,, \nn\\
	 \quad dR(\mu_1, \mu_2) > 0.1 \,, \quad k_T^\text{jet1} &< 50~\text{GeV} \,, \quad k_T^\text{jet2} < 		25~\text{GeV}\,,
\end{align}
where $\eta_i$ is the pseudo-rapidity of the muon from hard interaction $i$, $k_{T}^\text{lead}$ ($k_{T}^\text{subl}$) is the transverse momentum of the muon with the largest (second-largest) transverse momentum. $k_{T}^{\mu_3}$ is the transverse momentum of a third muon, $\slashed{E}_T$ the missing transverse momenta and $k_T^\text{jet1}$ and $k_T^\text{jet2}$ are the transverse momenta of the two hardest jets. $dR = \sqrt{ (\phi_1 - \phi_2)^2 + (\eta_1 - \eta_2)^2 }$, where $\phi_i$ is the azimuthal angle of $\mu_i$, measure the distance between the two muons. On top of this we apply b-tagging with efficiencies $75\%$ for $k_T^\text{jet} \in \{25-30 \}$~GeV, $80\%$ for $k_T^\text{jet} \in \{30-40 \}$~GeV and $85\%$ for $k_T^\text{jet} \in \{40-50\}$~GeV \cite{Aaboud:2018xwy,Sirunyan:2017ezt}. With these cuts, the cross sections are given in Tab.~\ref{tab:xsec}.
\begin{table}
\begin{tabular}{ c|c } 
 & $\sigma$ [fb] \\
 \hline
  DPS $W^+W^+$ & 0.51\\ 
 $W^+ W^+ jj$ & 0.03 \\ 
 $W^+Z$ & 1.77  \\ 
 $ZZ$ & 0.00 \\
 $t\bar{t}$ & 2.46
\end{tabular}
\caption{\label{tab:xsec}Signal and background cross sections in fb for the production of two positively charged muons, with the selection in \eqref{eq:sel}.}
\end{table} 

The two remaining relevant backgrounds are the $WZ$ and $t\bar{t}$, which we will now discuss separately. There are efficient techniques to suppress both of these backgrounds which however require dedicated work in connection with performing the measurements, and lie outside the scope of this letter. For $t\bar{t}$, demanding tight isolation of the produced muons is a very strong discriminant to separate prompt muons from muons produced by meson decays. Using vertex localization to further discriminate between these two cases can additionally aid the separation and improvements on b-tagging can also help \cite{Sirunyan:2018fpa,Aad:2016jkr,Chatrchyan:2012xi}.  A detailed examination of all these possibilities should be done by the experimental collaborations performing the actual measurement. Based on our investigations we will assume that this type of discrimination in combination with data driven subtractions, can reduce the top background to 1\% of the above cross section with only a minor impact on the signal. For example, we could reduce the $t\bar{t}$ background by more than 95\%, keeping more than 90\% of the signal through crude muon isolation requirements (limiting the scalar transverse momentum sum of particles in a cone around the muon). The $WZ$ is a relatively clean process theoretically and can be accurately calculated. Already today the theoretical calculations for the total cross section have been made with high precision \cite{Grazzini:2017ckn,Grazzini:2016swo}. This background can be effectively suppressed through methods of multivariate analysis. For instance, with a naive application of a forest of decision trees we could enhance the signal to background ratio to about 1, with a signal cross section around 0.3 fb. We will assume that a dedicated multivariate analysis will achieve a WZ-background suppression to a level of one third of the signal. We will investigate how sensitive the measurement is to these assumptions by including an additional uncertainty to the measurements.

The most promising variable for detection of spin correlations is the asymmetry between the rate at which the two muons are created in opposite ($\sigma^-=\sigma(\eta_1\eta_2 < 0)$) compared to same ($\sigma^+=\sigma(\eta_1\eta_2 > 0$) hemispheres of the detector
\begin{align}
	A & = \frac{\sigma^- - \sigma^+}{\sigma^- + \sigma^+} \,.
\end{align}
In the absence of inter-parton correlations, the two muons are produced in two independent hard scatterings and the asymmetry from DPS will be zero. Therefore a non-zero asymmetry in DPS would be a clear indication of parton correlations. In particular polarization, with its direct influence on the muon rapidities, can produce large values of this asymmetry. It can be generated also by correlations between the longitudinal momentum fractions of the two partons \cite{Ceccopieri:2017oqe,Gaunt:2010pi}, but this typically give lower (but not necessarily negligible) values of the asymmetry \cite{Cotogno:2018xxx}. This type of correlation can  add a smaller but positive effect, increasing the total value of the asymmetry.

\begin{figure}[t]
  \centering
   \subfloat[]{\includegraphics[width=0.4\textwidth]{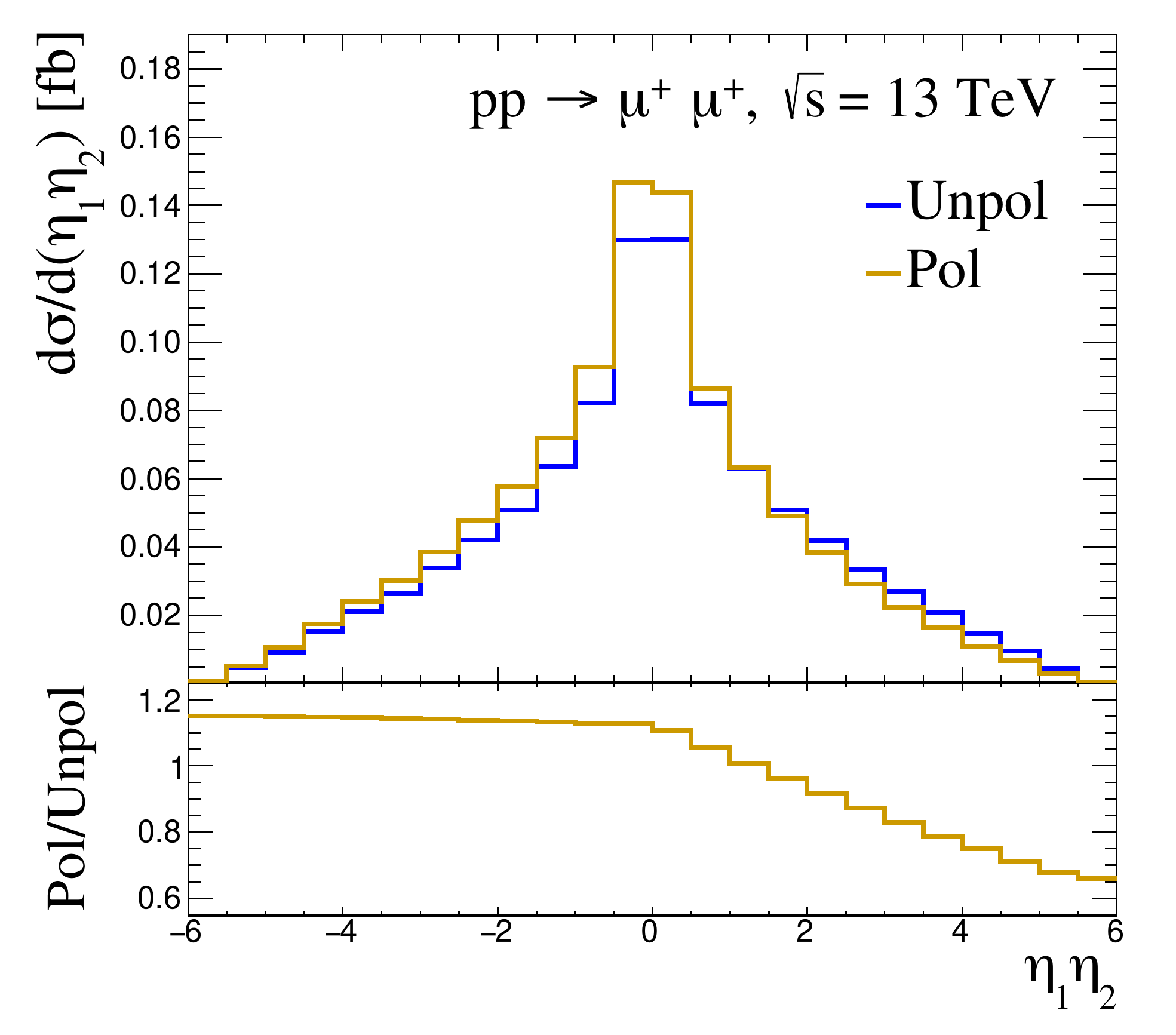}}
    \subfloat[]{\includegraphics[width=0.4\textwidth]{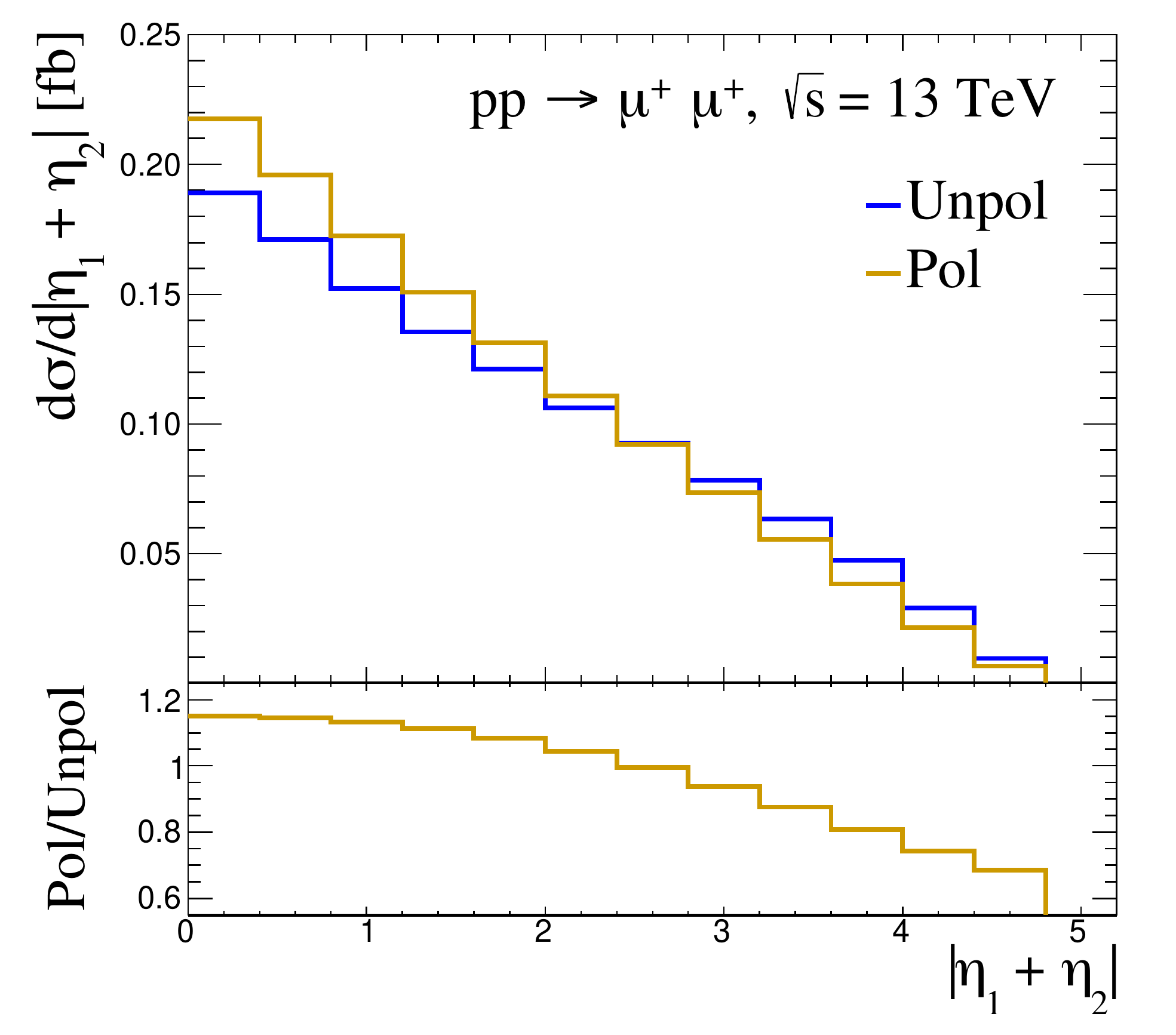}}
   \caption{Same-sign di-muon distributions in a) product of rapidities and b) rapidity sum, with (without) spin correlations in blue (yellow).}
   \label{fig:dists}
  \end{figure}

A less inclusive version of this asymmetry is the distribution in the product of the muon rapidities, shown in Fig.~\ref{fig:dists}a. 
This distribution is very sensitive to the spin correlations, and without correlations between the partons, it has to be symmetric around $\eta_1\eta_2 = 0$. In contrast, a non-symmetric $\eta_1\eta_2$ distribution can only be generated by parton correlations. The distribution as a function of the sum of the rapidities shows a clear dependence on the spin correlations and is an interesting observable to study DPS, as shown by Fig.~\ref{fig:dists}b. However, the slope of the distribution in the absolute sum of the rapidities is affected by the details of the properties of the single parton distributions used in the modeling of the unpolarized DPDs, and therefore it is a less clear indicator of partonic correlations.

The asymmetry calculated from the $\eta_1\eta_2$ distribution for our selection, which is a pure DPS sample, is shown in Tab.~\ref{tab:as}. 
\begin{table}
\begin{tabular}{ c|c|c|c } 
 $|\eta_i|$& $ > 0$  &  $ > 0.6$ &  $ > 1.2$ \\
 \hline
  $A$ & $0.07$ & $0.11$ & 0.16 \\
  $\sigma$ [fb] & $0.51$ & $0.29$ & $0.13$ 
\end{tabular}
\caption{\label{tab:as} Asymmetry and DPS cross section for different cuts on the pseudo-rapidities of the two muons.}
\end{table} 
An additional cut around central pseudorapidities $|\eta_i| > \eta_{\text{min}}$ increases the asymmetry, but naturally decreases the cross section. The values of the asymmetry depend directly on the model of the polarized DPDs. For example, taking $n=2$ in \eqref{eq:polmod} for all quarks and antiquarks results in a negative asymmetry. We want to emphasize that there are several ways in which this asymmetry can be further enhanced. For example by including smaller transverse muon momenta. The naive decision tree analysis performed to explore the potential power of a multivariate analysis to suppress the $WZ$ background naturally cause a small enhancement of the asymmetry. A full fledged multivariate analysis can simultaneously enhance the signal to background ratio and optimize the asymmetry. Based on our investigations in the simple decision tree analysis, we will assume that the 0.29 fb cross section with $A=0.11$ (as in \ref{tab:as}) can be reached with the $S/B = 3$. The contribution of the remaining $WZ$ background to the asymmetry can be subtracted by a precise theoretical calculation, on which we assume a further 10\% uncertainty. A detailed exploration on these possibilities is best done by the experimental collaborations and lies outside the scope of this letter. The extent to which our results are sensitive to this assumption will be investigated by varying the uncertainty related to the background. 

Next, we will examine how sensitive ATLAS and CMS will be to the asymmetry.  In order to do so, we estimate how many standard deviations away from zero a measurement of the   asymmetry will be. 
 We assume a Poissonian uncertainty on the number of DPS events with the two muons in the same/opposite hemispheres. We then create Gaussian distributions of the signal cross sections in the two hemispheres and use these to test how many standard deviations a measured asymmetry deviates from zero. 
In order to estimate the effect of our assumptions on the background, we add a parameter $b$ to the calculation of the uncertainty of the number of events per hemisphere (after subtraction)
~
$\Delta N = \sqrt{ [N_{WW}+b(N_{WZ} +N_{\text{top}})] + (b \Delta N^{\text{theo.}}_{WZ})^2} \,$
where $N_X$ is the number of events from process $X$, and $\Delta N^\text{theo.}_{WZ}$ is the theoretical uncertainty on the subtraction. The parameter $b$ is set to unity for the central predictions, and varied between 1/2 and 2 to give an uncertainty band indicating the sensitivity of our predictions on the totality of the background assumptions.

\begin{figure}[t]
   \includegraphics[width=0.45\textwidth]{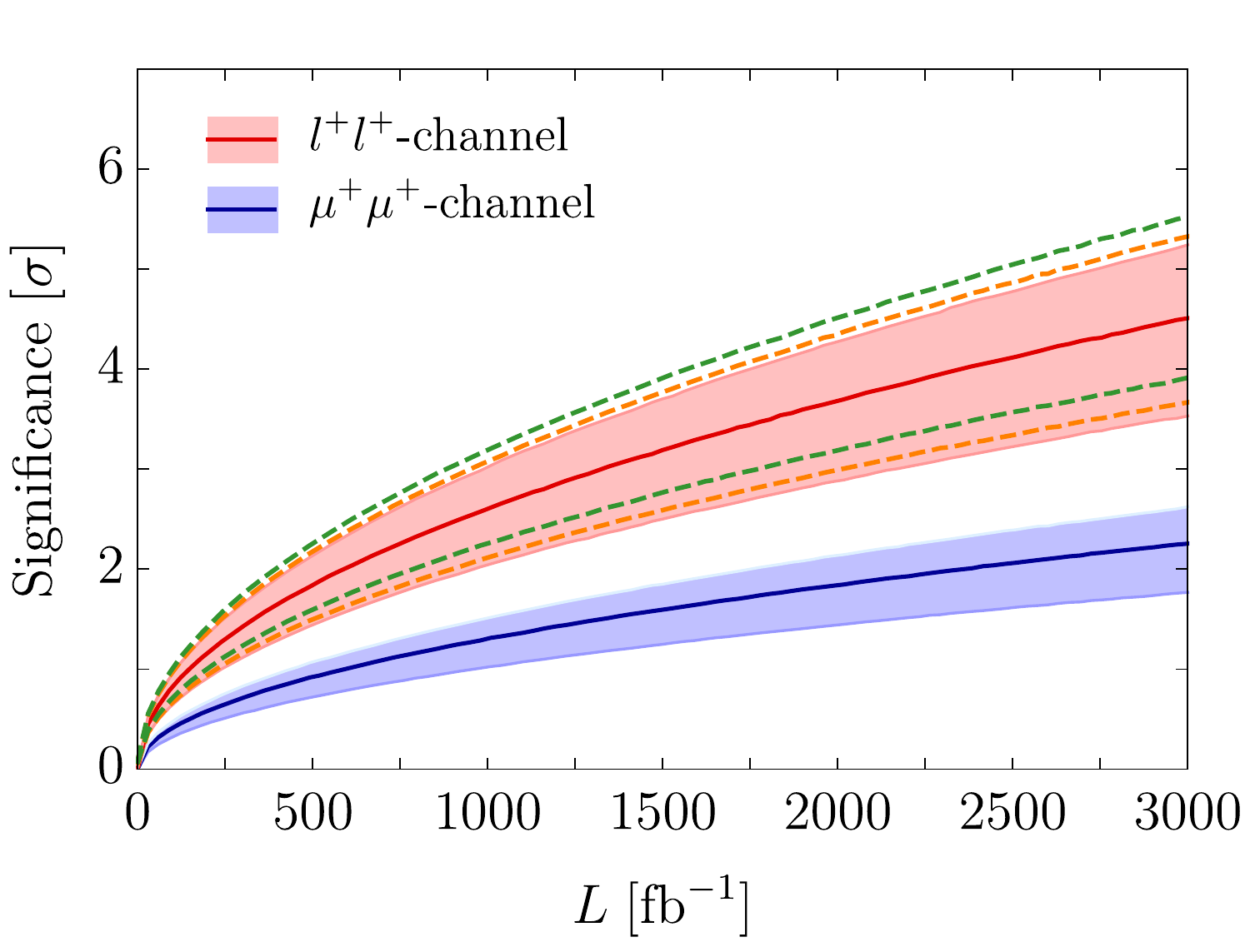}
   \caption{Estimate of the significance of a measured non-zero asymmetry of 0.11 for a signal cross section of 0.29 fb, as the distance in standard deviations of a measured asymmetry as in Tab.~\ref{tab:as} from zero.} Blue band for $\mu^+\mu^+$ final state only and red band including all positively charged combinations of two light leptons ($e$ and $\mu$). Dashed curves show the sensitivity of the central curve to changes of the asymmetry (modified by a factor of $1.2$  or $0.8$, orange dashed curves) and the magnitude of the DPS cross section (changed by a factor of $3/2$ or $3/4$, green dashed curves). 
   \label{fig:sigmas}
  \end{figure}

Fig.~\ref{fig:sigmas} shows the significance of a measurement as a function of the integrated luminosity. The central curve corresponds to $b=1$ while the band is obtained by the variation of $b$. 
With the $\mu^+\mu^+$ channel alone, a 3-sigma observation could be approached with the full high-luminosity LHC (HL-LHC) integrated luminosity of 3000 fb$^{-1}$. A detailed study of the backgrounds also for final states with electrons is beyond the scope of this article. If we assume that a similar sensitivity can be achieved for electrons as for muons, the DPS signals are equal and therefore effectively enhanced by a factor of 4 including all combinations of positively charged muons and electrons ($\mu^+\mu^+$, $e^+\mu^+$, and $e^+e^+$). This decreases the integrated luminosity necessary to measure the asymmetry, as shown in Fig.~\ref{fig:sigmas}. A 2-sigma hint is possible with around 400 fb$^{-1}$, a 3-sigma observation can be achieved with less than 1500 fb$^{-1}$ and an observation of a non-zero asymmetry approaching 5-sigma is reachable with the full 3000 fb$^{-1}$ \cite{Apollinari:2017cqg}. Figure~\ref{fig:sigmas} additionally demonstrates the effect on the significance coming from changes to the magnitude of the DPS cross section (multiplied by 3/2 or 3/4, green dashed curves) as well as the sensitivity to the value of the asymmetry (multiplied by a factor of 1.2 or  0.8). On top of this, combinations of measurements by CMS and ATLAS as well as additional possibilities with negatively charged leptons can further increase the sensitivity of the measurements. This means that there might be possible to see first indications of spin correlations before the start of the high-luminosity LHC. 

\section{Conclusions}
Quantum number correlations between two partons inside a hadron belong to the experimentally unexplored area of high energy physics. The measurement of same-sign muon pairs in W-boson pair production is one of the best candidates to detect these correlations, improving our knowledge on how closely the proton constituents are bound to one another. The left-handed nature of the coupling between quarks and W-bosons gives rise to a large dependence of the cross section on the polarized double parton distributions. We have presented a model of the polarized double parton distributions describing the spin correlations, with a large impact on the rate and distribution of the measured muons. We have demonstrated that the LHC has the potential to extract non-trivial information on spin-correlations between two quarks inside the proton through the measurement of the rate at which the leptons are produced in the same rather than opposite hemispheres. Due to their direct relation to the distribution of the final state leptons, the spin correlations are a prime suspect for the generation of a large asymmetry. The asymmetry is strongly model dependent, and even a measurement of a small or zero asymmetry would be of great value in constraining the range of possibilities for quantum correlations inside the proton. This asymmetry could be the first of a series of measurements at the LHC (HL-LHC), including additional observables in SSW and processes with one or two $W$'s replaced by other vector-bosons or hadrons, pinpointing how the quantum properties of quarks and gluons are connected to each other inside the proton.

\section*{Acknowledgements}
We thank Piet J. Mulders for useful discussions and Jonathan Gaunt and Lisa Zeune for valuable suggestions and feedback on the manuscript. SC acknowledges support from the European Research Council (ERC) under the program QWORK (contract no. 320389). TK acknowledges support from the Alexander von Humboldt Foundation.
 MM acknowledges support from the grant LTC17038 of the INTER-EXCELLENCE program at the Ministry of Education, Youth and Sports of the Czech Republic. 
\appendix

\bibliographystyle{JHEP}
\bibliography{KasBib}

\end{document}